\documentclass[12pt]{iopart}

\usepackage{latexsym}
\usepackage{graphicx}

\newcommand {\be}{\begin{equation}}
\newcommand {\ee}{\end{equation}}
\newcommand {\ba}{\begin{array}}
\newcommand {\ea}{\end{array}}
\newcommand {\bea}{\begin{eqnarray}}
\newcommand {\eea}{\end{eqnarray}}
\newcommand {\bean}{\begin{eqnarray*}}
\newcommand {\eean}{\end{eqnarray*}}

\newcommand {\eqref}[1]{{(\ref{#1})}}
\renewcommand{\etal}{\emph{et al.} \,}

\usepackage[english]{babel}
\usepackage[latin1]{inputenc}
\usepackage{times}

\begin{document}

\title{Post-Newtonian Initial Data with Waves: Progress in Evolution}

\author{B J Kelly$^{1,2}$, W Tichy$^3$, Y Zlochower$^4$, M Campanelli$^4$ and B Whiting$^5$}
\address{$^1$ CRESST \& Gravitational Astrophysics Laboratory, NASA/GSFC, 8800 Greenbelt Rd., Greenbelt, MD 20771, USA}
\address{$^2$ Dept. of Physics, University of Maryland, Baltimore County, 1000 Hilltop Circle, Baltimore, MD 21250, USA}
\ead{bernard.j.kelly@nasa.gov}
\address{$^3$ Department of Physics, Florida Atlantic University, Boca Raton, FL 33431, USA}
\address{$^4$ Center for Computational Relativity and Gravitation, School of Mathematical Sciences, Rochester Institute of Technology, Rochester, NY 14623, USA}
\address{$^5$ Institute for Fundamental Theory, Department of Physics, University of Florida, Gainesville, FL 32611, USA}

\begin{abstract}
In Kelly \etal \cite{Kelly:2007uc}, we presented new binary black-hole initial
data adapted to puncture evolutions in numerical relativity. This data satisfies
the constraint equations to 2.5 post-Newtonian order, and contains a
transverse-traceless ``wavy'' metric contribution, violating the standard assumption
of conformal flatness. We report on progress in evolving this data with a modern
moving-puncture implementation of the BSSN equations in several numerical codes.
We discuss the effect of the new metric terms on junk radiation and
continuity of physical radiation extracted.
\end{abstract}


\maketitle

\section{Introduction}
\label{sec:intro}

The astounding success of numerical relativity in simulating the merger of
comparable-mass black-hole binaries in recent years stemmed from a number of
numerical approaches to initial data, evolution formulations, gauge conditions,
and even grid structures. However, many active groups have converged on a
simple combination of methods called the ``moving puncture'' prescription
\cite{Campanelli:2005dd,Baker:2005vv}.

To initialise the numerical fields for a puncture evolution, most groups use
the puncture prescription of Brandt \& Br\"{u}gmann \cite{Brandt:1997tf}
with Bowen-York extrinsic curvature \cite{Bowen:1980yu}. In this scheme,
the three-metric $\gamma_{ij}$ is conformally flat:
\be
\gamma_{ij} = \psi^4 \eta_{ij}, \label{eq:gij_conformal}
\ee
and the conformal factor $\psi$ must satisfy the Hamiltonian constraint
\be
\Delta \psi + \frac{1}{8} K^{ab}K_{ab} \psi^{-7} = 0, \label{eq:CHam_conformal}
\ee
where the conformal extrinsic curvature $K_{ij}$ already satisfies the momentum
constraint for holes with arbitrary momentum and spin.
The zero-momentum constraint can be solved exactly to yield the
Brill-Lindquist solution for a pair of holes at a point of time-symmetry
\cite{Brill:1963yv}
\be
\psi = 1 + \frac{m_1}{2 |\vec{x}-\vec{x}_1|} + \frac{m_2}{2 |\vec{x}-\vec{x}_2|}. \label{psi_BL}
\ee
Here the $m_A$ are ``bare'' or ``puncture'' masses, residing at positions
$\vec{x}_A$ on the numerical grid. However, to solve Eq.~\eqref{eq:CHam_conformal}
in general requires numerical methods, and the infinities encountered at the puncture locations
are problematical. Brandt \& Br\"{u}gmann's insight was that the divergent parts
of $\psi$ could be formally factored out, leaving a well-behaved, simply connected
sheet on which to solve their modified constraint.

With the Brandt-Br\"{u}gmann prescription, only a single elliptic equation
has to be solved, and several fast solvers have been developed to perform this
operation to extremely high precision. The physical mass $M_A$ of each black hole
after solution will be greater than its puncture mass $m_A$.
 
\begin{figure*}[t]
\rotatebox{-90}{\includegraphics[clip,height=7.0in]{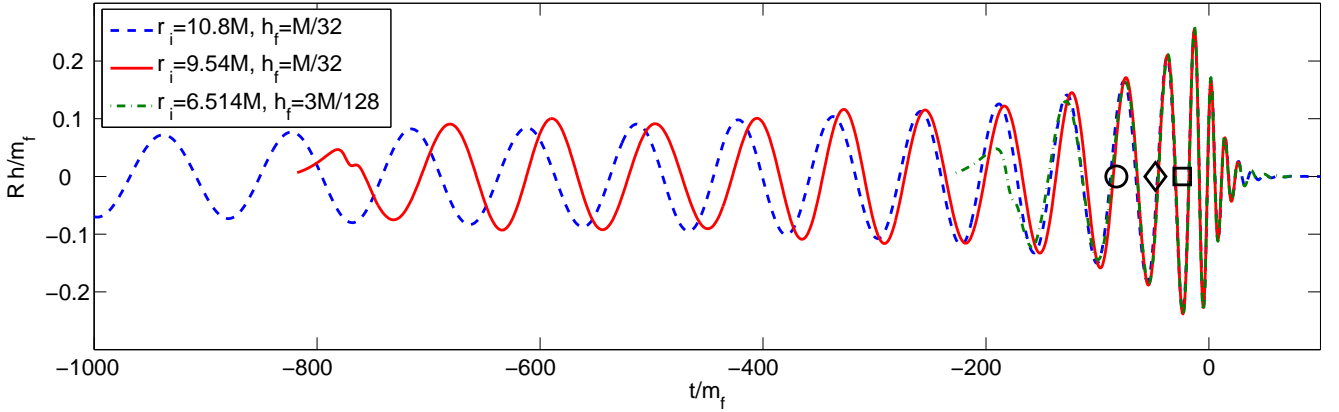}}
\caption{Strain waveforms for an equal-mass, zero-spin binary of total mass $M$, from
three initial separations $r_i$ ($h_{\rm f}$ indicates the resolution of the highest-
refinement region in the numerical simulation supplying each waveform). Here the time
axis is measured in units of the remnant mass $m_{\rm f} < M$ of the post-merger hole,
while the strain $h$ is multiplied by the extraction radius $R$ (again in units of
$m_{\rm f}$). Note that the waveforms in the $9.54M$ and $6.51M$ cases exhibit an initial
``glitch''. Adapted with permission from \cite{Baker:2006kr}.}
\label{fig:WFglitch}
\end{figure*}

However, the restriction of this data is that it is, by construction, conformally
flat. We know that the Kerr metric, the archetypal stationary solution of Einstein's
vacuum equations, is not conformally flat unless it has vanishing spin. It would seem
that requiring conformal flatness of an astrophysically realistic binary (which has
significant orbital angular momentum by construction) is unrealistic. In practice, when
evolving puncture binary data, we see early bursts of unphysical high-frequency radiation
propagate through the domain; see Fig.~\ref{fig:WFglitch}, reproduced from
\cite{Baker:2006kr}, as an example of this radiation at different initial separations.

\section{Post-Newtonian Metric in the ADM-TT Gauge}
\label{sec:admtt}

In the 1970s, Ohta \etal \cite{Ohta:1973qi,Ohta:1974uu,Ohta:1974kp} derived conditions for
the post-Newtonian metric of an $N$-particle system in the transverse-traceless ADM
(ADM-TT) gauge. The structure of this metric was given to 2.5 PN order by Sch\"{a}fer
\cite{Schaefer85}:
\be
\gamma_{ij} = \psi_{\rm PN}^4 \eta_{ij} + h^{TT}_{ij}\,,
\label{eq:gammaPN}
\ee
where the post-Newtonian conformal factor $\psi$ is expanded as
\be
\psi_{\rm PN} = 1 + \frac{1}{8}\left(  \frac{1}{c^2} \phi_{(2)} + \frac{1}{c^4} \phi_{(4)} + \cdots \right).
\label{eq:psiPN}
\ee
In this expression, $\phi_{(2)}$ alone yields the Brill-Lindquist potential term
\eqref{psi_BL} for two stationary, nonspinning particles, while the leading corrections
in $\phi_{(4)}$ depend explicitly on the separation of the particles, and on their
momenta. Note that the three-metric $\gamma_{ij}$ is no longer conformally flat, due to
the presence of a transverse-traceless term $h^{TT}_{ij}$. This satisfies an outgoing
wave condition:
\be
h^{TT}_{ij} = - \Box^{-1}_{ret} \delta^{TT\,kl}_{ij} \left[ \sum_{A=1}^N \, \frac{p_{Ak} \, p_{Al}}{m_A} \delta(x-x_A) + \frac{1}{4} \phi^{(2)}_{,k} \phi^{(2)}_{,l} \right].\label{eq:hTT_general}
\ee

The corresponding extrinsic curvature is derived from the conjugate post-Newtonian
three-momentum:
\be
\pi^{ij} = \psi_{\rm PN}^{-4} \left[ \frac{1}{c^3} \tilde{\pi}^{ij}_{(3)} + \frac{1}{c^5} \left( ( \phi_{(2)} \tilde{\pi}^{ij}_{(3)} )^{TT} + \frac{1}{2} \dot{h}^{TT}_{ij} \right)+ \cdots \right]\,.
\label{eq:piPN}
\ee
Explicit expressions were given for the terms $\phi_{(2)}$, $\phi_{(4)}$ and $\tilde{\pi}^{ij}_{(3)}$
by \cite{Schaefer85}, who also suggested a ``near zone'' approximation
for $h^{TT}_{ij}$, by splitting the retarded inverse d'Alembertian in \eqref{eq:hTT_general}
with an inverse Laplacian:
\bea
h^{TT}_{ij} &=& - [ {\Delta^{-1}} + (\Box^{-1}_{ret}  - \Delta^{-1} ) ] \delta^{TT\,kl}_{ij} \left[ \sum_{A=1}^N \, \frac{p_{Ak} \, p_{Al}}{m_A} \delta(x-x_A) + \frac{1}{4} \phi^{(2)}_{,k} \phi^{(2)}_{,l} \right]\\
            &=& {h^{TT\,(NZ)}_{i j}} + h^{TT\,(remainder)}_{i j} + O(v/c)^5.
\eea
The explicit form of this near-zone approximation was supplied by \cite{Jaranowski:1997ky}.

Tichy \etal \cite{Tichy:2002ec} adapted the ADM-TT-gauge 2.5PN results
to puncture initial data for NR. They established that although the ADM-TT metric was not
conformally flat, the behaviour of the metric near the black holes was dominated by the
conformal factor, and so fixed-puncture evolution methods should work as for standard puncture
data.

A few years later Kelly \etal \cite{Kelly:2007uc} completed the picture for nonspinning
binaries by determining the ``remainder'' TT term, $h^{TT\,(remainder)}_{i j}$ to 2PN order.
In the next section, we highlight some of the main properties of the complete solution.

\section{Summary of Global Properties of Solution}
\label{sec:properties}

\begin{figure}
\includegraphics[clip,width=6.0in]{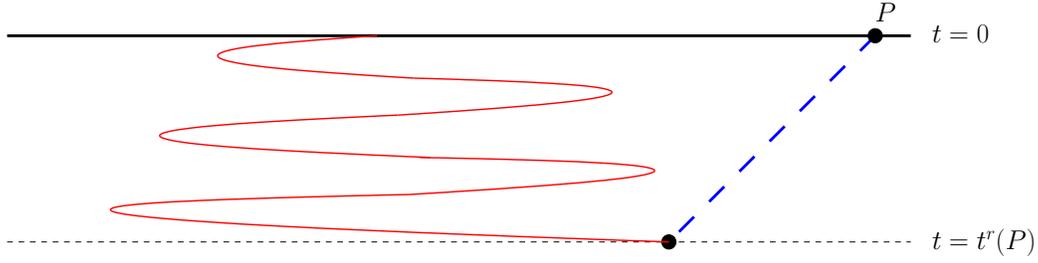}
\caption{Descriptive plot of the portion of the trajectory (thin curve) of particle 1 relevant
to the evaluation of $h^{TT}$ at field point $P$. $h^{TT\,(NZ)}_{i j}$ and $h^{TT\,(present)}_{i j}$ are
evaluated with the instantaneous positions and momenta at initial time $t=0$. $h^{TT\,(retarded)}_{i j}$
is evaluated using the instantaneous positions and momenta at the retarded time $t=t^r$ relative to $P$ (found by
travelling back along the light cone, indicated by a thick dashed line, until it intersects the particle's
trajectory). Finally, $h^{TT\,(interval)}_{i j}$ is evaluated by integrating along the trajectory segment
from $t=t^r$ to $t=0$.}
\label{fig:lightcone}
\end{figure}

While the work of \cite{Schaefer85} and \cite{Jaranowski:1997ky} applies to general systems
of particles, the ``remainder'' term presented in \cite{Kelly:2007uc} applies only to the
simplified situation of a binary system ($N=2$). In such a system, they determined that the structure 
of the remainder term divides into three segments, according to time of evaluation:
\be
h^{TT\,(remainder)}_{i j} = h^{TT\,(present)}_{i j} + h^{TT\,(retarded)}_{i j} + h^{TT\,(interval)}_{i j}.
\ee
For each field point where $h^{TT}_{i j}$ is to be evaluated, the ``present'' term is
evaluated using the particle positions and momenta at $t=0$, the time at which the simulation
will start. The ``retarded'' term is evaluated using positions and momenta at the retarded time
of each source particle relative to the field point. Finally, the ``interval'' term is an integral
over the particles' paths from the retarded time to the present. Figure~\ref{fig:lightcone}
illustrates this division.

Each of these segments, moreover, consists of a ``kinetic'' and a ``potential'' part, the
former depending on the particles' momenta, the latter on their relative positions.

In this expression, the present-time piece almost completely \emph{cancels} the near-zone
solution of \cite{Jaranowski:1997ky}: the kinetic terms cancel exactly, while the slightly
more involved potential terms are suppressed by three powers of the field distance $R$:
\be
h^{TT\,(NZ + present)}_{ij} = \frac{G^2\,m_1\,m_2\,r}{16\,{R^3}} \left\{ \cdots \right\} +  \mathcal{O}\left(\frac{1}{R^4}\right).
\ee

\begin{figure} 
\includegraphics[clip,width=5.0in]{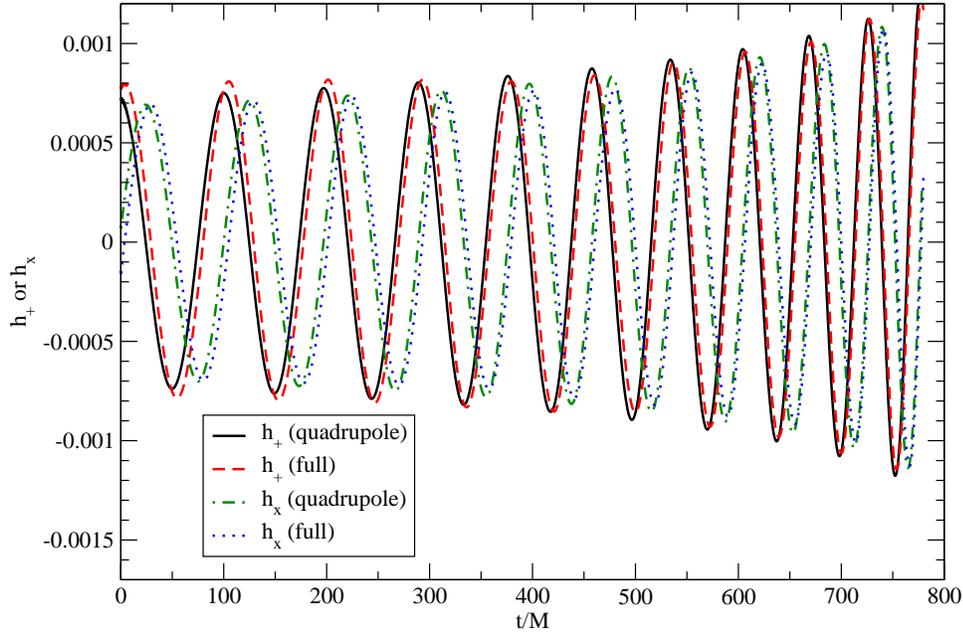}
\caption{$h^{TT\,(full)}$ versus quadrupole for an equal-mass binary, evaluated along the orbital axis.
Reproduced from \cite{Kelly:2007uc}.}
\label{fig:hTTfullVSquad}
\end{figure}

The retarded-time piece reduces to the well-known quadrupole solution for a nonspinning binary
as $r/R \rightarrow 0$. The most involved term, the interval piece, is too difficult to do in
generality, and must be integrated numerically. In Figure~\ref{fig:hTTfullVSquad}, we evaluate
the full solution $h^{TT}_{ij}$ over time, along the orbital axis of an equal-mass system, assuming
a simple inspiral. We see that the waveform is very close to the quadrupole solution.

\begin{figure}
\includegraphics[clip,width=4.0in]{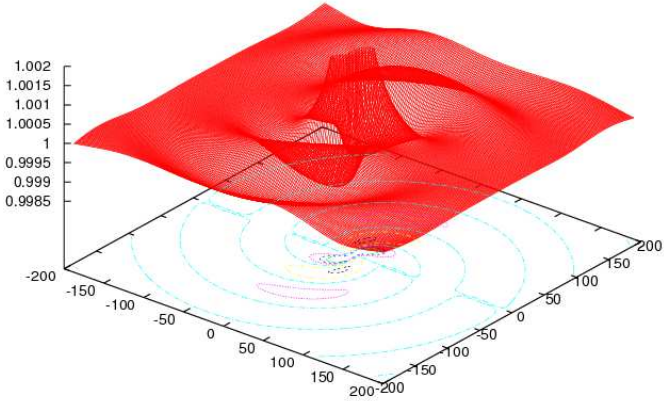}
\caption{Three-metric component $\gamma_{xx}$ in the $x$-$y$ plane with all terms for an equal-mass binary.}
\label{fig:gxxfull}
\end{figure}

Figure~\ref{fig:gxxfull} shows one of the three-metric components for
the full solution, with the characteristic quadrupole swirl we expect from an inspiralling binary.

\section{Encoding the Binary's Past Inspiral}
\label{sec:inspiral}

To evaluate the retarded-time and time-interval contributions, $h^{TT\,(retarded)}_{i j}$,
$h^{TT\,(interval)}_{i j}$, we need a model for the past history of the black holes.
Initially, in \cite{Kelly:2007uc}, we employed a hybrid procedure (here $M = M_1 + M_2$ is
the total mass of the system, while $\eta \equiv M_1 M_2/M^2$ is the \emph{symmetric mass ratio}):

\begin{enumerate}
\item obtain separation $r$ as a function of orbital frequency $\Omega$ to 2PN order by
      (numerically) inverting the ``puncture adapted'' relation from \cite{Tichy:2003qi}:
\be
M \Omega = \sqrt{ \frac{64 M^3 r^3}{(M + 2 r)^6} + \frac{1}{c^2} \frac{M^4 \eta}{r^4} + \frac{1}{c^4} \frac{M^5 (-5 \eta + 8 \eta^2)}{8 r^5} };
\ee
\item obtain transverse momentum $p$ as a function of $\Omega$ to 2PN order from
      Sch\"{a}fer \& Wex \cite{Schaefer93}:
\be
p(\Omega) = M \eta \left[ (M \Omega)^{1/3} + \frac{(15 - \eta)}{6} (M \Omega) + \frac{(441 - 324 \eta - \eta^2)}{72} (M \Omega)^{5/3} \right];
\ee
\item obtain orbital phase $\Phi$ (and hence frequency $\Omega \equiv d\Phi/dt$) as a
      function of time to 2PN order using the explicit relation from \cite{Blanchet:2006LR}:
\be
\Phi(t) = \Phi(t_c) - \frac{1}{\eta}\,\left[ \Theta^{5/8} + \left( \frac{3715}{8064} + \frac{55}{96} \eta \right)\,\Theta^{3/8} - \frac{3 \pi}{4} \, \Theta^{1/4} + \cdots \right],
\ee
where $\Theta \equiv \eta \, (t_c - t) / 5\,M$, and $t_c$ is the (nominal) merger time.
\end{enumerate}

This method has several drawbacks. For one thing, it is quite limited in post-Newtonian
order. For another, the components are in inconsistent gauges. Finally, we have no
prescription for an instantaneous radial momentum, necessary for low-eccentricity
inspiral.

A conceptually simpler approach is to get all the needed information from a single source.
Following recent practice in initial parameters for numerical evolutions of punctures
\cite{Husa:2007rh}, we can evolve the binary system inspiral through Hamiltonian evolution
of the PN equations of motion. This has been shown by Husa \etal \cite{Husa:2007rh} to result
in extremely low eccentricity, at least for nonspinning data. For more generic spinning
binaries, the situation is more complicated, but promising; see, for instance, \cite{Campanelli:2008nk}.

Although we need much more information for the new data, the Hamiltonian evolution method
should perform just as well as for simple punctures: the puncture positions and momenta
required are all from earlier times, and hence larger separations with lower velocities,
where post-Newtonian methods are guaranteed to work. The only drawback is that potentially a
lot of data must be stored about the past history of the binary to calculate the retarded and
interval terms: and the larger the numerical grid, the further back in time we must reach.

\section{Numerics}
\label{sec:numerics}

In the geometric units ($G = c = 1$) commonly used for vacuum numerical relativity, time, length,
and mass can be scaled by a single number. For this purpose we use $M = M_1 + M_2$, the total
mass of the binary system.

The numerical implementation of the wavy PN data has taken place in three independent codes,
the Cactus-based \texttt{LazEv} code \cite{Zlochower:2005bj,Campanelli:2005dd,Lousto:2007rj}, the
\texttt{BAM} code \cite{Bruegmann:2003aw,Bruegmann:2006at,Husa:2007hp}, and the Paramesh-based
\texttt{Hahndol} code \cite{Imbiriba:2004tp,vanMeter:2006vi}. Some of the code was auto-generated using a
\texttt{Mathematica} script supplied by Gerhard Sch\"{a}fer. For simplicity, the
time-derivatives of $h^{TT}_{ij}$ appearing in the 2.5PN extrinsic curvature calculation \eqref{eq:piPN}
are carried out using simple second-order-accurate centred differencing (this is easily
accurate enough, as the time spacing used is much smaller than the spatial discretisation of the
numerical grid).

To calculate the two retarded times for each field point (one for each black hole), we use a Newton
solver. Interval terms are then calculated by integrating from these retarded times to the present
using Romberg integration.

\begin{figure}
\rotatebox{-90}{\includegraphics[clip,height=3.0in]{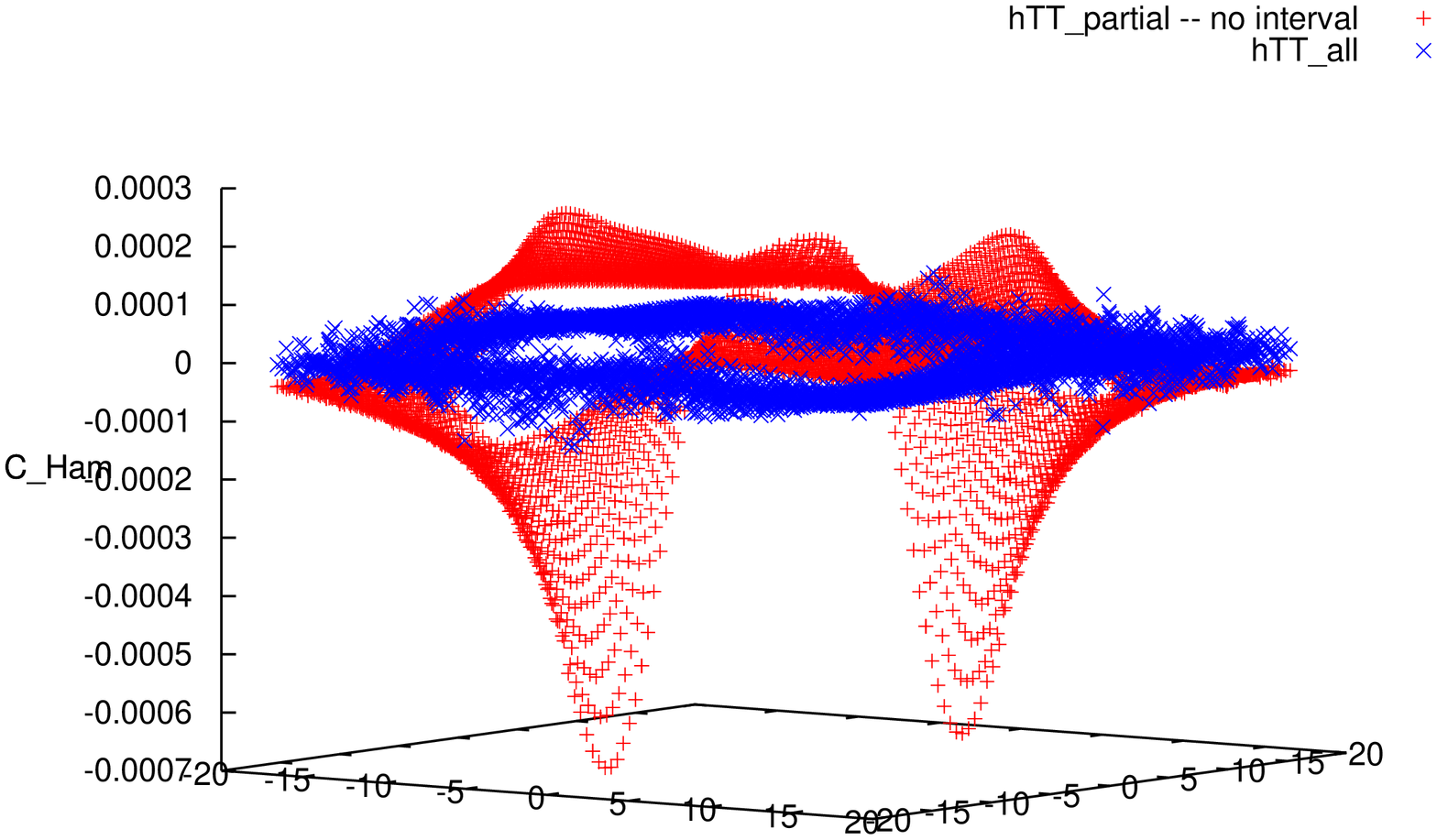}}
\rotatebox{-90}{\includegraphics[clip,height=3.0in]{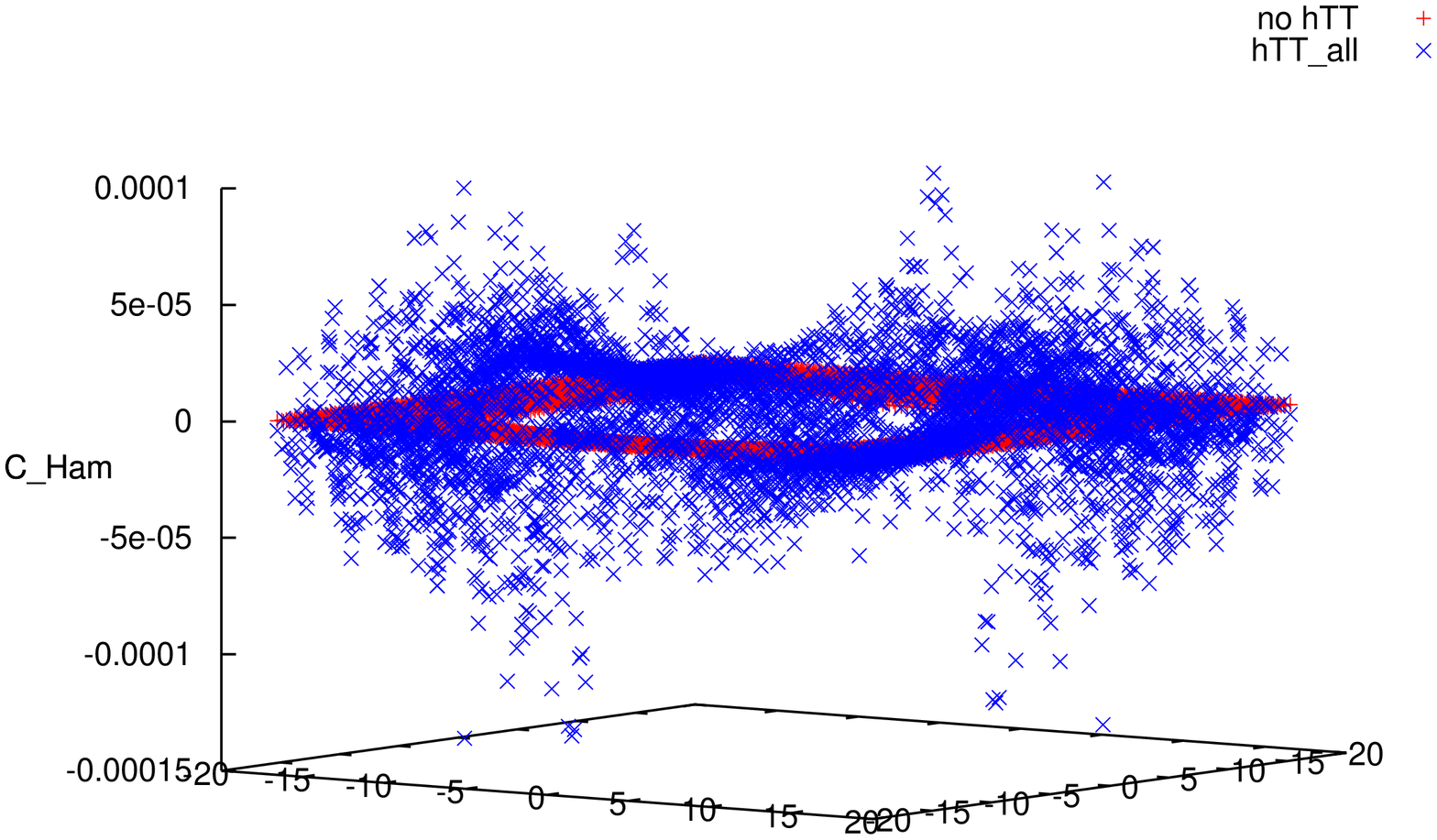}}
\caption{Left panel: Hamiltonian constraint in the orbital ($x$-$y$) plane, with (dashed) and without
(solid) $h^{TT\,(interval)}_{i j}$. Right panel: Hamiltonian constraint in the same plane with full
$h^{TT}$ (dashed) and with no $h^{TT}$ (solid).}
\label{fig:C_Ham_violations}
\end{figure}

Before discussing the evolution of the data, we note two numerical properties of the initial data.
The first is that, as expected, the Hamiltonian constraint violation for the complete solution is
better than for a partial solution: the left panel of Fig.~\ref{fig:C_Ham_violations} demonstrates
this with the Hamiltonian constraint evaluated for the complete solution with and without the
``interval'' term.

On the other hand, it seems that leaving out the $h^{TT}$ terms \emph{entirely} yields even lower
constraint violations, as seen in the right panel of Fig.~\ref{fig:C_Ham_violations}. The reason
for this is unclear, but may have to do with the greatly increased number of numerical evaluations
in the $h^{TT}$ terms. Given the observed behaviour of the black-hole (horizon) masses described in
the next section, it seems that the bulk of the constraint violation ``falls back'' into the hole, with
little escaping to the field zone to pollute the waveforms. We discuss the residual constraint violation
further in Sec.~\ref{sec:future}.

\section{Early Evolution Results}
\label{sec:results}

Initial evolutions have been carried out for equal-mass nonspinning binaries, with initial separations
between $6M$ and $10M$. All evolutions show a combination of desirable and undesirable effects,
which we illustrate with figures from $10M$-separation evolutions at low central resolution ($3M/128$
near the punctures) with the Goddard \texttt{Hahndol} code. We will often compare with the results of
a more traditional moving puncture evolution of Bowen-York puncture data, with initial separation of $11M$,
using parameters from \cite{Husa:2007rh}. For this data, we solved the Hamiltonian constraint numerically
using Marcus Ansorg's \texttt{TwoPunctures} spectral code \cite{Ansorg:2004ds}. In both cases, we tracked
apparent horizons before and after merger with Jonathan Thornburg's \texttt{AHFinderDirect} code
\cite{Thornburg:2003sf}.

\subsection{Eccentricity and Horizon Masses}

\begin{figure}
\includegraphics[clip,height=3.0in]{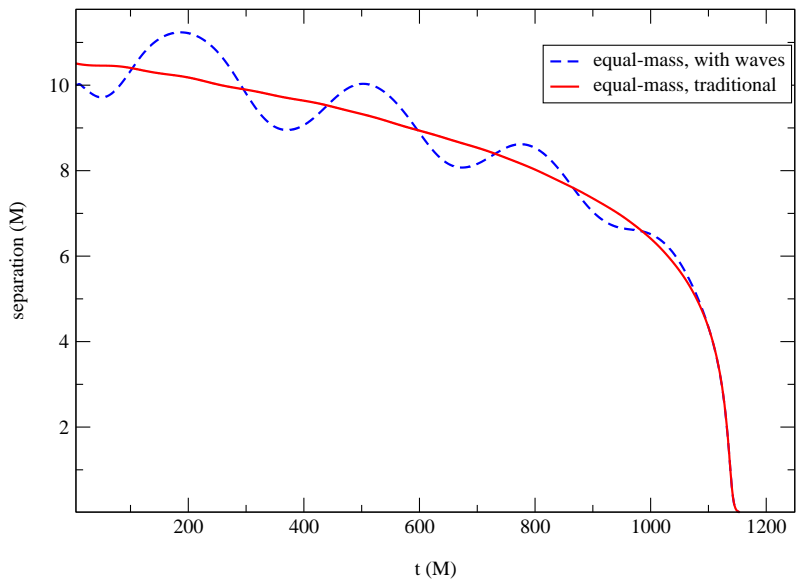}
\caption{Coordinate separation of traditional ``solved" data (solid) and wavy data (dashed). The wavy
data exhibits almost 10\% eccentricity initially.}
\label{fig:D10_eccentricity}
\end{figure}

The first thing we note is the presence of strong eccentricity in the puncture tracks of the holes --
see Fig.~\ref{fig:D10_eccentricity}. This eccentricity appears to be around 10\%, far greater than that
of the traditional evolution, and persists until around $100M$ before merger.

\begin{figure}
\includegraphics[clip,height=3.0in]{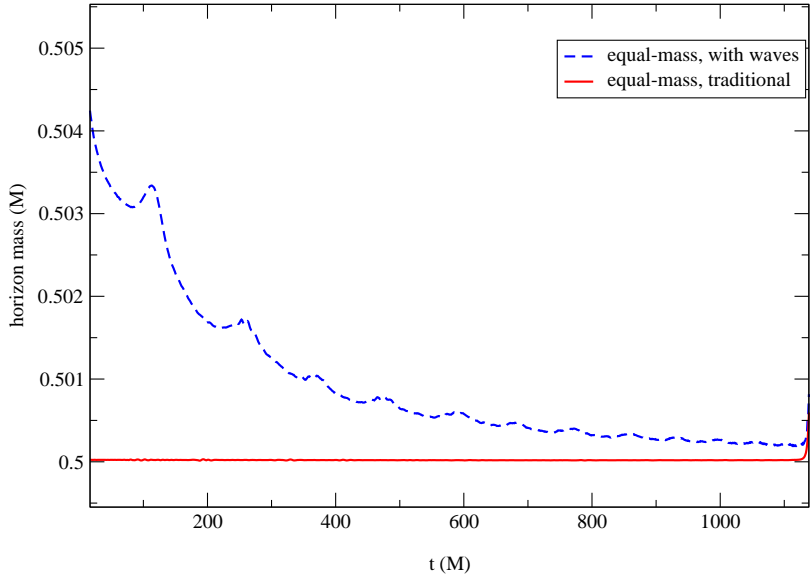}
\caption{Pre-merger AH mass for black hole 1 for traditional data (solid) and wavy data (dashed).
Over the course of the inspiral, the wavy data horizon mass drops by around 1\%. Variation in
the mass for the traditional data is negligible by comparison.}
\label{fig:D10_MAH_evolve}
\end{figure}

A related phenomenon may be seen in the evolution of apparent horizon masses for the pre-merger
binary -- see Fig.~\ref{fig:D10_MAH_evolve}. The apparent horizon mass $M_{\rm AH}$ of a hole is
related to the area of the apparent horizon. We locate the former numerically using the
\texttt{AHFinderDirect} code \cite{Thornburg:2003sf}. We calculate the proper area $A_{\rm AH}$
of this 2-surface, and derive from this the \emph{irreducible mass}
$M_{\rm irr} \equiv \sqrt{A_{\rm AH}/16\pi}$. The horizon mass is then related to $M_{\rm irr}$ by
inverting the Christodoulou formula \cite{Christodoulou:1970wf}:
\be
M_{\rm AH}^2 = \frac{2 M_{\rm irr}^2}{\hat{a}^2} \left[ 1 - \sqrt{1 - \hat{a}^2} \right], \label{eq:Mirr2MAH}
\ee
where $\hat{a}\equiv |\vec{S}|/M^2$ is the dimensionless spin of the hole. Note that for zero-spin
holes, the horizon mass $M_{\rm AH}$ is identical to the irreducible mass $M_{\rm irr}$.

In Fig.~\ref{fig:D10_MAH_evolve} we see that while the standard horizon masses
quickly settle down to a stable value, varying insignificantly over the $1100M$ of pre-merger
evolution, the unsolved wavy data masses decay at a much slower rate, and with a periodic
saw-tooth feature over time. If this is not merely an artifact of the horizon-finding algorithm
or implementation, then the holes are losing mass steadily over the course of the inspiral.
As discussed in \cite{Bode:2009fq}, this could be associated with the nature of the residual
constraint violation on the initial time-slice. This
would lead to a considerable ambiguity in what the ``correct'' horizon mass is. The initial momenta
(as well as the entire past history in the $h^{TT}$ terms) depend sensitively on the hole's mass;
an incorrect mass might result in considerable eccentricity.

\subsection{Waveforms}
\label{ssec:waveforms}

\begin{figure}
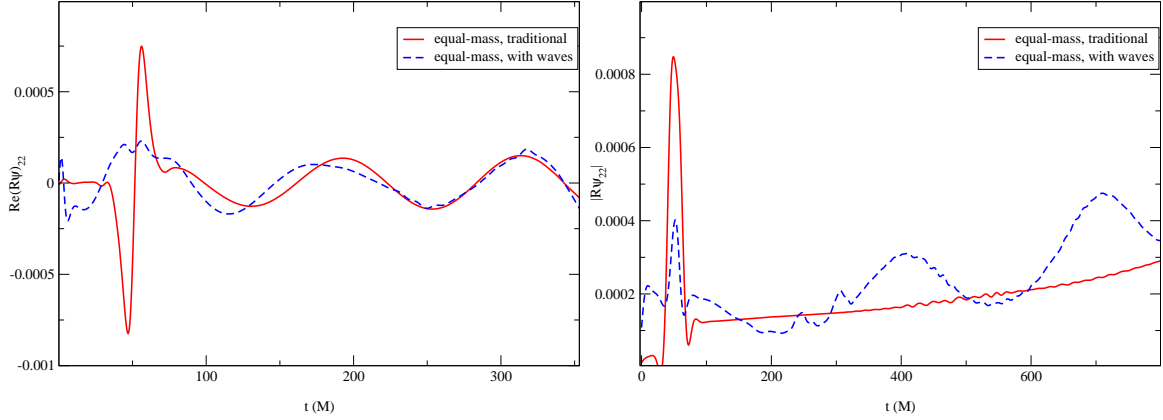

\includegraphics[clip,width=3.0in]{D10_RePsi4_comp.eps}
\includegraphics[clip,width=3.0in]{D10_AmpPsi4_comp.eps}
\caption{(2,2) mode of $R\psi_4$ from $t = 0$ for standard puncture evolution (solid) and
wavy data (dashed), as extracted at $R=45M$. The left panel shows the real part of the mode,
which already has a non-zero value at $t=0$, and a smaller non-physical pulse around $t=45M$,
compared with the traditional moving-puncture waveform. The right panel shows the amplitude of
complex mode. The high eccentricity visible in the (gauge-dependent) puncture tracks has a
clear effect here also, compared with the low-eccentricity traditional evolution.}
\label{fig:D10_Psi4}
\end{figure}

The main quantity of interest from these evolutions must be the gravitational
radiation extracted. In the left panel of Fig.~\ref{fig:D10_Psi4} we show the real part
of the dominant $(\ell=2,m=2)$ waveform mode $R\psi_{4}$, as extracted on the coordinate
sphere $R = 45M$.
We note that indeed the physical waveform is present from $t=0$, with less junk radiation
than is present in the standard solved data.

Unfortunately a poor choice in the structure of the numerical grid in the radiation zone
led to noisy extraction for both traditional and new data.
Note that a high-accuracy waveform would be extracted at (or extrapolated to)
$R \rightarrow \infty$; however, this was not possible with the numerical grid used for these initial
evolutions. Coupled with the relatively low resolution used, we estimate waveform amplitude
errors of up to $\sim$ 8\% at the peak; thus our numerical results are qualitative in nature
at this point. Higher-accuracy extraction and analysis methods will
be appropriate in future evolutions, when the mass and eccentricity issues described above
have been resolved.

\subsection{Final State}
\label{ssec:final_state}

\begin{figure}
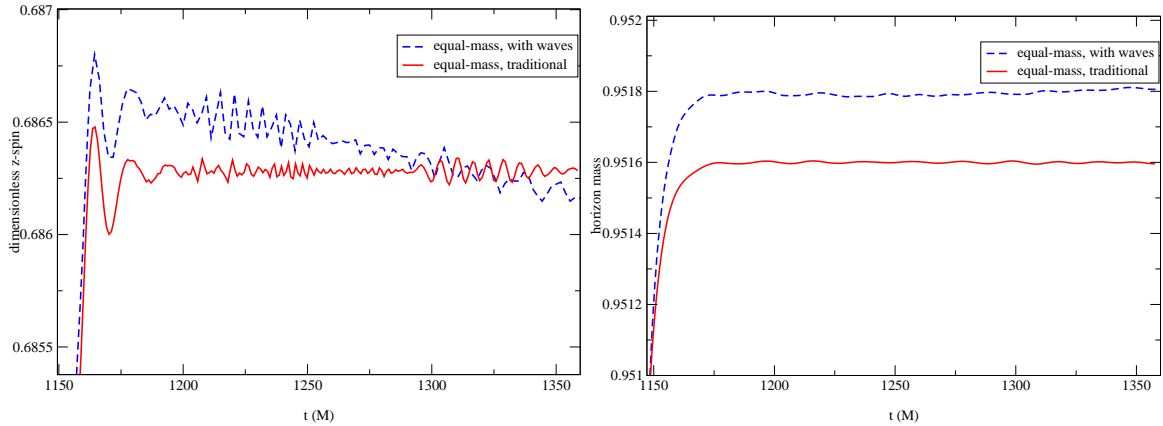

\includegraphics[clip,width=3.0in]{AH3_spin_comp.eps}
\includegraphics[clip,width=3.0in]{AH3_mass_comp.eps}
\caption{Left panel: post-merger $z$-spin $\hat{a} \equiv S_z/M^2$ for standard data (solid) and wavy data (dashed).
Right panel: Post-merger horizon mass for standard data (solid) and wavy data (dashed),
using Eq.~\eqref{eq:Mirr2MAH}, and assuming $\hat{a} = 0.68625$.}
\label{fig:D10_merged_spin_mass}
\end{figure}

The final state of the post-merger black hole, which we analyse using the \texttt{AHFinderDirect}
code \cite{Thornburg:2003sf}, is qualitatively similar to that of the standard solved data. The
final spin, as estimated by values of the Coulomb scalar on the surface of the apparent horizon
\cite{alignedIRS}, is around $\hat{a} \equiv S_z/M^2 \approx 0.68625$. Using \eqref{eq:Mirr2MAH},
we estimate the total horizon mass $M_{\rm AH}$ of the remnant hole.
Figure~\ref{fig:D10_merged_spin_mass} shows $\hat{a}$ and $M_{\rm AH}$ for the standard and 
wavy merger remnants.

\section{Open Questions \& Future Work}
\label{sec:future}

From these early evolutions, we see that the the new wavy PN data appears to achieve at
least some of our goals: it does evolve stably in the moving puncture recipe, without any
special tweaks in gauge or evolution equations. Moreover, the early waveforms do indeed
show more reasonable start-up behaviour than those of a standard puncture evolution, with
a physically reasonable non-zero value and a diminished nonphysical pulse.

Nevertheless, some aspects of the numerics are not satisfactory, most notably the high
orbital eccentricity and the slowly settling horizon masses of the pre-merger holes. These
two phenomena may be simply related, as our experience in working with standard puncture
data has shown us that small errors in tuning masses will lead to eccentricity. It is
conceivable that the residual Hamiltonian constraint violation of the wavy data causes
this mass defect; similar effects have been studied recently in \cite{Bode:2009fq}.

Resolving the mass issue may necessitate the introduction of a numerical elliptic solver
to remove the residual constraint violation. To do this completely is not straightforward,
as the ADM-TT gauge used here will change the form of the Hamiltonian constraint; moreover,
the momentum constraint will in general need to be solved too.

Looking beyond these issues, we may consider the introduction of spin to our data. Though
explicit post-Newtonian solutions of the constraint equations with spin are not yet available,
we note that the leading-order momentum contributions to the conformal extrinsic curvature
are just the Bowen-York momentum terms. It is conceivable that the leading-order spin
contributions are given by the corresponding Bowen-York terms also.

Finally, we note that we have not addressed the initial conditions of the lapse function and
shift vector. It is known that in standard puncture evolutions with the popular ``1+log'' slicing
conditions, these settle down at late times to a ``trumpet'' form \cite{Hannam:2008sg,Immerman:2009ns}.
Until these late-time forms can be incorporated into the full initial data, we cannot expect
to eliminate gauge pulses in our waveforms.

\ack

We would like to thank Bruno Mundim and Hiroyuki Nakano for useful comments during the preparation of
this manuscript.

We acknowledge support from NASA grant 06-BEFS06-19. BJK was supported in part
by an appointment to the NASA Postdoctoral Program at the Goddard Space Flight Center, administered by Oak
Ridge Associated Universities through a contract with NASA.

\section*{References}

%

\end{document}